\documentclass[aps,prd,10pt,nofootinbib,OliveGeqsecnum,showpacs,showkeys,superscriptaddress,preprintnumbers,altaffilletter,floatfix,twocolumn]{revtex4-2}

\usepackage{amsmath}
\usepackage{multirow}
\usepackage{hyperref}
\usepackage{soul}
\usepackage{graphicx}
\usepackage{dcolumn}
\usepackage{amssymb}
\usepackage{amsfonts}
\usepackage{amsbsy}
\usepackage{color}
\usepackage{rotating}
\usepackage{overpic}
\usepackage{ifxetex}
\usepackage{lipsum}
\usepackage{float}
\usepackage{bbold}
\usepackage{enumerate}
\usepackage{dsfont}
\usepackage{mathrsfs}
\usepackage{mathtools}
\usepackage{tensor}
\usepackage{caption}
\usepackage{subcaption}
\usepackage{ragged2e}
\ifxetex
   \usepackage{polyglossia}
   \setdefaultlanguage{english}
   \usepackage{fontspec}
\else
   \usepackage[english]{babel}
   \usepackage[utf8]{inputenc}
\fi

\definecolor{amaranth}{rgb}{0.9, 0.17, 0.31}
\definecolor{palatinateblue}{rgb}{0.15, 0.23, 0.89}
\definecolor{brightpink}{rgb}{1.0, 0.0, 0.5}

\hypersetup{ linktoc=all,
    colorlinks, linkcolor={palatinateblue},
    citecolor={brightpink}, urlcolor={amaranth}
}

\graphicspath{ {images/} }

\newcommand{\abs}[1]{\left\lvert #1\right\rvert}

\begin{document}

\title{Cosmological perturbations in a generalised axion-like dark energy model}

\author{Carlos G. Boiza}
\email{carlos.garciab@ehu.eus}
\affiliation{Department of Physics \& EHU Quantum Center, University of the Basque Country UPV/EHU, P.O. Box 644, 48080 Bilbao, Spain}
\author{Mariam Bouhmadi-López}
\email{mariam.bouhmadi@ehu.eus}
\affiliation{IKERBASQUE, Basque Foundation for Science, 48011, Bilbao, Spain}
\affiliation{Department of Physics  \& EHU Quantum Center, University of the Basque Country UPV/EHU, P.O. Box 644, 48080 Bilbao, Spain}

\begin{abstract} 

We analyse the cosmological evolution of a generalised axion-like field that drives the late-time acceleration of the Universe. This model can exhibit tracking behaviour, which alleviates the coincidence problem. The cosmological perturbations are carried within a multi-fluid approach where the scalar field is described by a non-adiabatic fluid, i.e., one whose speed of sound in the rest frame differs from the adiabatic one. The cosmological perturbations are solved since the radiation-dominated epoch, imposing initial adiabatic conditions for matter, radiation and the dark energy component, for modes well outside the Hubble horizon in the past. We analyse the homogeneous curvature perturbation, gravitational potential and dark energy perturbations in this model, as well as the matter power spectrum and f$\sigma_8$. We discuss which parameters of the model are more favoured observationally. 

\end{abstract}

\maketitle

\renewcommand{\tocname}{Index}


\section{Introduction}\label{intro}

In recent years, experiments have detected and confirmed a late-time accelerated expansion of the Universe \cite{SupernovaSearchTeam:1998fmf,SupernovaCosmologyProject:1998vns,Bahcall:1999xn}. The nature of this acceleration remains unknown. The simplest way to include such an accelerated expansion in a cosmological model is through the introduction of a cosmological constant, leading to the $\Lambda$CDM model. Although this model successfully explains most observations, it presents some unresolved issues. On the one hand, at a fundamental level, the cosmological constant introduces a new energy scale, $\rho_v\approx10^{-47}\,\textrm{GeV}^{\,4}$, which is very small compared to other known scales in particle physics \cite{Weinberg:1988cp,Sahni:1999gb}. Additionally, the fact that the amounts of matter and dark energy are comparable today, i.e., $\Omega_{m0}\sim\Omega_{\textrm{de}0}$, cannot be explained naturally within this model without fine-tuning the initial conditions. This is known as the coincidence problem \cite{Carroll:2000fy,Padmanabhan:2002ji}. On the other hand, recent observations indicate a tension in determining the current expansion rate $H_0$ of the Universe. Low-redshift measurements favour higher values of $H_0$ compared to those obtained through high-redshift measurements \cite{DiValentino:2021izs,Poulin:2018cxd,Kamionkowski:2022pkx, Vagnozzi:2023nrq}. A similar tension arises when determining the value of $S_8$ \cite{DiValentino:2020vvd,Perivolaropoulos:2021jda,Abdalla:2022yfr, Nunes:2021ipq}. There is also the so-called neutrino tension: cosmological constraints on the sum of neutrino masses are lower than those obtained through particle physics experiments \cite{Lesgourgues:2006nd,Hannestad:2006zg,Hannestad:2010kz,Wong:2011ip,Lesgourgues:2012uu,daFonseca:2023ury, Jiang:2024viw}. For all these reasons, models that go beyond the standard $\Lambda$CDM cosmological model have been proposed and studied in recent years, from perfect fluids \cite{Kamenshchik:2001cp,Nojiri:2005sx,Albarran:2016mdu} to modified theories of gravity \cite{CANTATA:2021asi}.

In this work, we focus on a model in which dark energy is described by a scalar field minimally coupled to gravity: a quintessence model. These types of models have proven useful in alleviating the coincidence problem through scaling solutions \cite{Wetterich:1987fm,Ferreira:1997hj} and tracking behaviour \cite{Zlatev:1998tr,Steinhardt:1999nw}. Specifically, we focus on the generalised axion-like dark energy model, motivated in \cite{Boiza:2024azh}. In this model, the scalar potential is given by $V(\phi)=\Lambda^4\left[1-\cos{(\phi/\eta)}\right]^{-n}$, where $\Lambda$, $\eta$ and $n$ are free positive parameters, which motivated us to name it generalised axion-like, since the same potential for $n<0$ naturally appears in axionic models in particle physics (see \cite{Ng:2000di,Hammer:2020dqp,Marsh:2015xka,Chakraborty:2021vcr} for details on axions applied to cosmology). The potential for $n<0$ has been previously analysed in the literature: it has been used to alleviate the coincidence problem through the string axiverse \cite{Kamionkowski:2014zda,Emami:2016mrt}, in early dark energy models to address the Hubble tension \cite{Poulin:2018cxd,Kamionkowski:2022pkx}, in Natural Inflation models \cite{Freese:1990rb}, in primordial black hole formation mechanisms \cite{Khlopov:2004sc} and also in axionic dark matter models \cite{Khlopov:1999tm,Hui:2016ltb}. Cosmological simulations based on ultralight axions have been performed in \cite{Schive:2014dra,Schive:2014hza,Mocz:2019uyd}. In our case, we assume $n>0$. In \cite{Boiza:2024azh}, we demonstrated how the potential is capable of generating tracking behaviour, allowing us to alleviate the coincidence problem. We also showed how the transition from matter to dark energy can be favoured in this model compared to others. In \cite{Hossain:2023lxs}, the authors studied the differences between the generalised axion-like model and an inverse power-law potential and concluded that the former is able to describe the Universe evolution in a more satisfactory way, as the transition from matter to dark energy occurs more rapidly and the equation of state today deviates less from $-1$.

To study structure formation, one must go beyond the background analysis conducted in \cite{Boiza:2024azh}. Perturbation theory provides a mathematical framework for analysing the inhomogeneities of the Universe and deducing observable quantities such as the matter power spectrum and the distribution of $f\sigma_8$, thus allowing us to compare the model expected results with observations related to the clustering of matter. Cosmological perturbations have been extensively studied in the literature, including those of a scalar field in quintessence models \cite{Mukhanov:1990me,Ma:1995ey,Brandenberger:2003vk,Malik:2008im,baumann,Malquarti:2002iu,Bassett:2005xm,Wang:2009azb}. However, we will show that in this work, it will suffice to adopt a multi-fluid picture, as used in \cite{Albarran:2016mdu}, since the cases considered do not involve oscillatory regimes.

The paper is organized as follows. In Sec. \ref{model}, we will introduce the generalised axion-like model and review its most important aspects, which we have already analysed in depth in \cite{Boiza:2024azh}, such as its tracking behaviour. In Sec. \ref{cosmpert}, we will introduce the framework for cosmological perturbations that we will use. In Subsec. \ref{frame}, we will review the general aspects of perturbations in a quintessence model and explain why the multi-fluid picture used in \cite{Albarran:2016mdu} will be sufficient for the cases studied here. In Subsec. \ref{admodandcurvpert}, we will introduce the adiabatic mode first studied by Weinberg in \cite{Weinberg:2003sw} along with the comoving and homogeneous curvature perturbations. In Subsec. \ref{inicond}, we will discuss the initial conditions chosen to numerically solve the perturbation equations in our model. In Sec. \ref{numres}, we will present and discuss the numerical results obtained. We will examine the behaviour of the homogeneous curvature perturbation, the gravitational potential and the dark energy perturbation in Subsecs. \ref{curvpert}, \ref{gravpot} and \ref{depert}, respectively. In Sec. \ref{cosobs}, we will study how structure formation operates in our model, analysing the matter power spectrum and the $f\sigma_8$ distribution. Finally, in Sec. \ref{conclusions}, we will conclude with some key aspects of our work.

\section{The model}\label{model}

In this section, we review the generalised axion-like model and the key aspects of its dynamics. We conducted a detailed analysis of it in \cite{Boiza:2024azh}. The model was proposed to explain the late-time acceleration of the Universe through a scalar field minimally coupled to gravity. It is a quintessence model whose action is given by

\begin{equation}\label{scalaraction}
   S=\int d^4x\sqrt{-g}\left(\frac{R}{2k^2}+L_m+L_r+L_{\phi}\right),
\end{equation}
where $k^2=8\pi G=1/M_P^2$ (we use natural units $\hbar=c=1$ throughout this work and $M_P\approx2.435\times10^{18}\,\textrm{GeV}$ is the reduced Planck mass), $R$ is the scalar curvature, $L_m$ is the Lagrangian corresponding to dark and baryonic matter,  $L_r$ is the Lagrangian corresponding to radiation and $L_{\phi}$ is the Lagrangian of a minimally coupled scalar field $\phi$:

\begin{equation}\label{canonicallagrangian}
   L_{\phi}=-\frac{1}{2}g^{\mu\nu}\partial_{\mu}\phi\partial_{\nu}\phi-V(\phi),
\end{equation}
where $V(\phi)$ is the potential of the scalar field and is given in this model by

\begin{equation}\label{axionscalarpotential}
   V(\phi)=\Lambda^4\left[1-\cos{(\phi/\eta)}\right]^{-n},
\end{equation}
where $\Lambda$, $\eta$ and $n$ are free positive parameters of the theory. In \cite{Boiza:2024azh}, we studied the generalised axion-like model in a homogeneous and isotropic spatially flat universe described by the spatially flat FLRW metric:

\begin{equation}\label{rwmetric}
   ds^2=-dt^2+a^2(t)d\bar{x}^2.
\end{equation}
Therefore, the equation of motion of the scalar field reads

\begin{equation}\label{sfeq}
   \ddot{\phi}+3H\dot{\phi}+V_{,\phi}=0,
\end{equation}
where a dot stands for derivative with respect to the cosmic time $t$ and $V_{,\phi}$ is the derivative of $V$ with respect to $\phi$. The Friedmann equation is given by

\begin{equation}\label{friedmanneq}
   H^2=\frac{k^2}{3}\left[\rho_m+\rho_r+\frac{1}{2}\dot{\phi}^2+V(\phi)\right],
\end{equation}
where $\rho_m$ and $\rho_r$ are the energy densities of dark and baryonic matter and radiation, respectively. Note that the last two terms of \eqref{friedmanneq} can be identified as the energy density of the scalar field: $\rho_{\phi}=\dot{\phi}^2/2+V(\phi)$. For completeness, we also introduce the pressure of the scalar field: $p_{\phi}=\dot{\phi}^2/2-V(\phi)$. Another useful quantity that we will use in the perturbation equations (see next section) is the fractional energy density which, for each component $A$, reads as $\Omega_A=\rho_A/\rho$, where $\rho$ is the total energy density: $\rho=\sum_A\rho_A$.

In \cite{Boiza:2024azh}, we expanded the potential \eqref{axionscalarpotential} around its minimum and found a non-vanishing constant term which plays the role of an effective cosmological constant:

\begin{equation}\label{minimumexpansion}
   V(\phi)\approx\frac{\Lambda_{\textrm{eff}}}{k^2}+\frac{1}{2}m^2(\phi-\pi\eta)^2,
\end{equation}
where $\Lambda_{\textrm{eff}}=k^2\Lambda^4/2^n$ and $m^2=n\Lambda^4/(2^{n+1}\eta^2)$. The non-vanishing constant term appeared in the dynamical system as a late-time attractor completely dominated by the potential term of the scalar field, allowing for a late-time de Sitter universe in which the equation of state is $-1$. In this way, we were able to explain the late-time acceleration of the Universe. In addition to the effective cosmological constant, the potential also includes a mass term fully specified by the parameters of the theory.

A crucial aspect of the model is that it exhibits tracking behaviour. In \cite{Boiza:2024azh}, we argued that for the model to be distinguishable from $\Lambda$CDM, we should initially be far from the minimum of the potential, $\phi_i/\eta\ll1$, as we will next explain.

We start by defining the parameter $\lambda$ as \cite{Copeland:1997et,Steinhardt:1999nw,delaMacorra:1999ff,Ng:2001hs}

\begin{equation}\label{lambdadef}
   \lambda=-\frac{V_{,\phi}}{kV}.
\end{equation}
We can write $\lambda$ as a function of $\phi$ in our model as \cite{Boiza:2024azh}

\begin{equation}\label{lambdaphimodel}
   \lambda(\phi)=\frac{n}{k\eta}\frac{\sin{(\phi/\eta)}}{1-\cos{(\phi/\eta)}}.
\end{equation}
In terms of $\lambda$, the initial condition $\phi_i/\eta\ll1$ is translated into an initial value that satisfies $\lambda_i\gg1$. We next define the $\Gamma$ function as \cite{Steinhardt:1999nw,delaMacorra:1999ff,Ng:2001hs}

\begin{equation}\label{gammadef}
   \Gamma=\frac{VV_{,\phi\phi}}{V_{,\phi}^2}.
\end{equation}
In our model, $\Gamma$ is given by \cite{Boiza:2024azh}

\begin{equation}\label{gammamodel}
   \Gamma=1+\frac{1}{2n}+\frac{n}{2k^2\eta^2\lambda^2}.
\end{equation}
Since $n>0$, the $\Gamma$ function satisfies $\Gamma>1$. Additionally, in the regime where $\lambda\gg1$, when we are far from the minimum, $\Gamma$ is well approximated by a constant expression that only depends on the parameter $n$:

\begin{equation}\label{gammamodelapprox}
   \Gamma\approx1+\frac{1}{2n}.
\end{equation}
Both conditions, $\Gamma>1$ and $\Gamma\approx const.$, are the necessary conditions that guarantee physically interesting tracking behaviour with $w_{\phi}<w$, where $w$ is the equation of state of the dominant background fluid, as postulated in \cite{Steinhardt:1999nw}. Assuming that the dominant background fluid in the tracking regime is matter, with an equation of state $w=0$, we showed in \cite{Boiza:2024azh} that the equation of state of the scalar field in this regime is well approximated by an expression that depends only on the parameter $n$:

\begin{equation}\label{sfeosgammamodel}
   w_{\phi}\approx-\frac{1}{1+n}.
\end{equation}

In addition to the tracking behaviour, we showed in \cite{Boiza:2024azh} how the generalised axion-like model is capable of favouring a rapid transition to a dark energy-dominated universe at late-time. In a tracking regime dominated by the background fluid, the density parameter of the scalar field scales with time as $\Omega_{\phi}\propto t^{P}$, where $P$ is expressed in terms of $\Gamma$ as \cite{Steinhardt:1999nw}

\begin{equation}\label{growgamma}
   P=\frac{4(\Gamma-1)}{1+2(\Gamma-1)}.
\end{equation}
If $\Gamma$ grows rapidly in the final stage of the tracking regime, a rapid transition from matter to dark energy dominance will occur. In our model, after the tracking regime, the scalar field approaches the minimum of the potential and the convergence toward the de Sitter attractor with $\lambda=0$ occurs much faster than in the case of an inverse power-law potential \cite{Hossain:2023lxs}. $\Gamma$ grows rapidly and, with it, $\Omega_{\phi}$. Other models with tracking behaviour, such as inverse power-law potentials $\phi^{-n}$, where $\Gamma$ remains constant, do not exhibit this effect. In \cite{Hossain:2023lxs}, the authors noted this difference between the inverse power-law potentials and the generalised axion-like model, observing that the equation of state of the latter is closer to $-1$ nowadays, making it more favourable.

It should be noted that, although the techniques used to study the dynamical system in \cite{Boiza:2024azh} are general and can be applied to any quintessence model, the analysed model has two main features that are not found in a general quintessence model. The first of these is the tracking behaviour. This type of behaviour can be found in other types of potentials, such as inverse power-law potentials. However, the model analysed here has a second feature that differentiates it from the former, as first noted in \cite{Hossain:2023lxs}. The transition from the matter-dominated era to the dark energy-dominated era occurs more rapidly, which makes the scalar field's equation of state today closer to $-1$ and, therefore, more favourable  observationally than the inverse power-law potential. All these features are achieved by simply adding a positive term of order $\mathcal{O}(1/\lambda^2)$ to the characteristic $\Gamma$ function of the inverse power-law potential.

We recall the parameter choices made in \cite{Boiza:2024azh}. For a critical density today of $\rho_c\sim10^{-47}\,\textrm{GeV}^{\,4}$ ($H_0\sim10^{-33}\,\textrm{eV}$), assuming that the Universe is dominated by the potential part of the scalar field near the minimum, we deduce from \eqref{minimumexpansion} that $\Lambda_{\textrm{eff}}/k^2\sim10^{-47}\,\textrm{GeV}^{\,4}$. Therefore, $\Lambda^4/2^n$ is fixed: $\Lambda^4/2^n\sim10^{-47}\,\textrm{GeV}^{\,4}$. In addition, for the scalar field to be rolling close to the minimum by today, we need $m\sim H_0\sim10^{-33}\,\textrm{eV}$, which imposes $\eta^2/n\sim10^{37}\,\textrm{GeV}^{\,2}$. For $n\sim1$, both conditions are satisfied if $\Lambda\sim10^{-3}\,\textrm{eV}$ and $\eta\sim M_P$. As in \cite{Boiza:2024azh}, we first fix the values of $\eta$ and $n$ and then calculate the value of $\Lambda$ imposing that today the density parameters of matter and radiation are approximately given, respectively, by $\Omega_{m0}=0.3$ and $\Omega_{r0}=8\times10^{-5}$ ($\Omega_{\phi0}$ is related to the other density parameters through $\Omega_{\phi0}=1-\Omega_{m0}-\Omega_{r0}$). Following \cite{Boiza:2024azh}, we fix the parameter $\eta$ to $M_P$ and consider two different values of the parameter $n$: $n=1$ and $n=0.1$. This parameter choice ensures that we are free of oscillations at the bottom of the potential and that we can use the multi-fluid picture when studying the perturbations (see next section). We find $\Lambda\approx2.177\times10^{-3}\,\textrm{eV}$ and $\Lambda\approx2.102\times10^{-3}\,\textrm{eV}$ for the two cases $n=1$ and $n=0.1$, respectively. 

To conclude this section, we note how the selection of parameters influences the evolution of the scalar field and, consequently, the evolution of the Universe. In \cite{Boiza:2024azh}, we saw that in the case of $n=1$, there was a suppression of $\Omega_m$ due to the difference between $w_{\phi}\approx-1/2$ in the tracking regime (cf. \eqref{sfeosgammamodel}) and $w_v=-1$, corresponding to a cosmological constant. For the case of $n=0.1$, the difference is smaller and the suppression is relaxed ($w_{\phi}\approx-10/11$ in this case). In fact, in the limit $n\rightarrow0$, we recover a flat potential in \eqref{axionscalarpotential} and the model becomes equivalent to a cosmological constant ($w_{\phi}\rightarrow-1$ in the tracking regime when $n\rightarrow0$). In the following sections, we will explore the effect of this suppression, which depends on the value of $n$, on the cosmological perturbations and structure formation. 

\section{Cosmological perturbations}\label{cosmpert}

Cosmological perturbations have been extensively analysed in the literature \cite{Mukhanov:1990me,Ma:1995ey,Brandenberger:2003vk,Malik:2008im,baumann}. In this section, we study the linear regime of the perturbations in our model in the Newtonian gauge. We focus on the scalar perturbations and follow an approach similar to the one used in \cite{Albarran:2016mdu}.

\subsection{Framework}\label{frame}

The perturbed FLRW metric in the Newtonian gauge reads

\begin{equation}\label{perturbedmetric}
   ds^2=a^2(\eta)[-(1+2\Phi(\eta,x^i))d\eta^2+(1-2\Psi(\eta,x^i))d\bar{x}^2],
\end{equation}
where $\eta$ is the comoving time, $d\eta=dt/a$, and $\Phi(\eta,x^i)$ and $\Psi(\eta,x^i)$ are the scalar metric perturbations, which depend on the comoving time and on the spatial coordinates $x^i$ and coincide with the Bardeen potentials \cite{Bardeen:1980kt}. In the absence of anisotropies, the potentials are equal: $\Phi=\Psi$. From now on, we assume that the fluid perturbations do not introduce any anisotropy and that the equality is thus satisfied. 

In this work, we follow the multi-fluid approach used in \cite{Albarran:2016mdu}. In this framework, for each fluid $A$, the evolution equations for the fractional energy density perturbation, $\delta_A=\delta\rho_A/\rho_A$, and the peculiar velocity potential, $v_A$, are given by \footnote{For each component $A$, we can write the diagonal components of the perturbed energy-momentum tensor as $\delta T^0_{A0}=-\delta\rho_A$ and $\delta T^i_{Aj}=\delta p_A\delta^i_ j$, where $\delta\rho_A$ and $\delta p_A$ are, respectively, the perturbation of the energy density and the perturbation of the pressure of the component $A$. The non-diagonal components will be given by $\delta T^0_{Ai}=(p_A+\rho_A)\partial_iv_A$, where $v_A$ is the peculiar velocity potential. We note that, since we have assumed that we are free of anisotropies, there will be no purely spatial non-diagonal components. The evolution equations \eqref{originalperturbationeq1} and \eqref{originalperturbationeq2} are easily deduced from the perturbed energy-momentum conservation equations for each component $A$: $\nabla_{\mu}\delta T^{\mu}_{A\nu}+\delta\Gamma^{\mu}_{\mu\alpha}T^{\alpha}_{A\nu}-\delta\Gamma^{\alpha}_{\mu\nu}T^{\mu}_{A\alpha}=0$, where $\nabla_{\mu}$ is constructed from the Christoffel symbol of the background FLRW metric and $\delta\Gamma^{\alpha}_{\mu\nu}$ is the perturbation of the Christoffel symbol.}

\begin{equation}\label{originalperturbationeq1}
\begin{split}
   \delta'_A=\,\,&3(w_A-c^2_{sA})\delta_A+3(1+w_A)\Psi'\\
   &+(1+w_A)\left[9\mathcal{H}(c^2_{sA}-c^2_{aA})-\frac{\nabla^2}{\mathcal{H}}\right]v_A,
\end{split}
\end{equation}

\begin{equation}\label{originalperturbationeq2}
   v'_A=(3c^2_{sA}-1)v_A-\frac{c^2_{sA}}{\mathcal{H}(1+w_A)}\delta_A-\frac{\Psi}{\mathcal{H}},
\end{equation}
where a prime stands for a derivative with respect to $\log{(a/a_0)}$, with $a_0$ being the current value of the scale factor and $\mathcal{H}$ is the conformal Hubble parameter, $\mathcal{H}=aH$. We have introduced the effective and the adiabatic speeds of sound for each fluid $A$. They respectively read as

\begin{equation}\label{speedsofsound}
   c^2_{sA}=\frac{\delta p_A}{\delta\rho_A}\bigg|_{\textrm{r.f.}}, \quad c^2_{aA}=\frac{p'_A}{\rho'_A}.
\end{equation}
The effective speed of sound is defined in the rest frame (r.f.) of the fluid.

In our model, radiation and matter are described by perfect fluids with constant equations of state of $w_r=1/3$ and $w_m=0$, respectively. Both fluids are adiabatic and, since $c^2_{sA}=c^2_{aA}=w_A$ for $A=r,m$\,, \eqref{originalperturbationeq1} and \eqref{originalperturbationeq2} simplify to

\begin{equation}\label{deltaradperturbationeq}
   \delta'_r=\frac{4}{3}\left(\frac{k^2}{\mathcal{H}}v_r+3\Psi'\right), \quad v'_r=-\frac{1}{\mathcal{H}}\left(\frac{\delta_r}{4}+\Psi\right),
\end{equation}

\begin{equation}\label{deltamatperturbationeq}
   \delta'_m=\frac{k^2}{\mathcal{H}}v_m+3\Psi', \quad v'_m=-\left(v_m+\frac{\Psi}{\mathcal{H}}\right).
\end{equation}
Note that we have performed a Fourier transformation. The equations for each mode are obtained after the substitution $\nabla^2\rightarrow-k^2$ in \eqref{originalperturbationeq1} and \eqref{originalperturbationeq2}. From now on, we will work with the Fourier components.

The scalar field perturbations are more complicated but can be accommodated into this multi-fluid picture (see, for example, \cite{Malquarti:2002iu,Bassett:2005xm,Wang:2009azb}) by making certain assumptions that are satisfied in our model. Perturbing the energy-momentum tensor of the scalar field \footnote{The energy-momentum tensor of the scalar field is given by $T^{\phi}_{\mu\nu}=\partial_{\mu}\phi\partial_{\nu}\phi-\frac{1}{2}g_{\mu\nu}g^{\alpha\beta}\partial_{\alpha}\phi\partial_{\beta}\phi-g_{\mu\nu}V(\phi)$. The expressions \eqref{scalardensityperturbation}, \eqref{scalarpressureperturbation} and \eqref{scalarvelocityperturbation} are obtained by perturbing the tensor and writing it in the same form as in the case of a fluid: $\delta T^0_{\phi0}=-\delta\rho_{\phi}$, $\delta T^i_{\phi j}=\delta p_{\phi}\delta^i_ j$ and $\delta T^0_{\phi i}=(p_{\phi}+\rho_{\phi})\partial_iv_{\phi}$.}, we find the expressions for the energy density perturbation, the pressure perturbation and the peculiar velocity potential of the scalar field in terms of $\delta\phi$:

\begin{equation}\label{scalardensityperturbation}
   \delta\rho_{\phi}=\frac{\mathcal{H}^2}{a^2}\phi'(\delta\phi'-\phi'\Psi)+V_{,\phi}\delta\phi,
\end{equation}

\begin{equation}\label{scalarpressureperturbation}
   \delta p_{\phi}=\frac{\mathcal{H}^2}{a^2}\phi'(\delta\phi'-\phi'\Psi)-V_{,\phi}\delta\phi,
\end{equation}

\begin{equation}\label{scalarvelocityperturbation}
   v_{\phi}=-\frac{\delta\phi}{\mathcal{H}\phi'}.
\end{equation}

The peculiar velocity \eqref{scalarvelocityperturbation} is ill-defined if $\phi'=0$. In that case, the kinetic part of the scalar field is null and the potential term completely dominates. The equation of state of the scalar field becomes $w_{\phi}=-1$. Note that in the freezing regime, when the scalar field remains almost constant, the equation of state is close to $-1$ and the potential term dominates over the kinetic part. Nevertheless, this is an approximate regime and, in fact, there is a small evolution that saves us from this problem. The perturbations can be calculated using the variables $\delta_{\phi}=\delta\rho_{\phi}/\rho_{\phi}$ and $v_{\phi}$, as supported by the numerical results obtained in the work and presented in the next section. Another potentially problematic scenario arises if the field oscillates around the minimum of the potential. In that case, each time the field changes the direction of the oscillation, the derivative of the field becomes null, $\phi'=0$, and the equation of state takes the value $w_{\phi}=-1$. This is not an approximate regime and it is really problematic. This framework fails to reproduce finite results and an approach different from the multi-fluid one is needed. In this work, we discuss cases with $\eta=M_P$ in which the field is rolling close to the minimum nowadays and asymptotically reaches the minimum, but it does not cross it, at least at the times computed. Therefore, the fluid approach is sufficient for us. Perturbations in an oscillatory regime ($\eta<M_P$) will be analysed in a work in preparation \cite{cgbmariam}.

In order to define the evolution equations \eqref{originalperturbationeq1} and \eqref{originalperturbationeq2} for the scalar field ($A=\phi$), we need to find the expressions for $w_{\phi}$, $c^2_{s\phi}$ and $c^2_{a\phi}$. The equation of state is given by $w_{\phi}=p_{\phi}/\rho_{\phi}$ and can be written in terms of background quantities as

\begin{equation}\label{fieldeos}
   w_{\phi}=\frac{\dot{\phi}^2/2-V(\phi)}{\dot{\phi}^2/2+V(\phi)}.
\end{equation}
From \eqref{scalardensityperturbation}, \eqref{scalarpressureperturbation} and \eqref{scalarvelocityperturbation}, we find the relation

\begin{equation}\label{scalardensitypressurerelation}
   \delta p_{\phi}=\delta\rho_{\phi}+2\mathcal{H}\phi'V_{,\phi}v_{\phi}.
\end{equation}
We find that the second term on the right hand side is proportional to $v_{\phi}$. Therefore, the scalar field does not behave like an adiabatic fluid. The pressure perturbation of a fluid $A$ can be separated into its adiabatic part and its non-adiabatic part as \cite{Bean:2003fb,Valiviita:2008iv}

\begin{equation}\label{adandnonadpressure}
   \delta p_{A}=c^2_{aA}\delta\rho_{A}+\delta p_{naA}.
\end{equation}
Following \cite{Valiviita:2008iv}, and also used in \cite{Albarran:2016mdu}, we can express the non-adiabatic part in the Newtonian gauge as

\begin{equation}\label{nonadpressure}
   \delta p_{naA}=(c^2_{sA}-c^2_{aA})(\delta\rho_{A}+\mathcal{H}\rho'_Av_A).
\end{equation}
Introducing the last expression into \eqref{adandnonadpressure}, we finally obtain \footnote{We have used the background conservation equation, given by $\rho'_A+3(1+w_A)\rho_A=0$, to eliminate $\rho'_A$ in \eqref{finaladandnonadpressure}.}

\begin{equation}\label{finaladandnonadpressure}
   \delta p_{A}=c^2_{sA}\delta\rho_{A}-3\mathcal{H}(c^2_{sA}-c^2_{aA})(1+w_A)\rho_Av_A.
\end{equation}
Comparing \eqref{finaladandnonadpressure} for a general fluid with the relation \eqref{scalardensitypressurerelation} valid for the scalar field, we find the expressions for $c^2_{s\phi}$ and $c^2_{a\phi}$ in terms of the background components:

\begin{equation}\label{scalarspeeds}
   c^2_{s\phi}=1, \quad c^2_{a\phi}=1+\frac{2a^2V_{,\phi}}{3\mathcal{H}^2\phi'}.
\end{equation}
Note that, again, the adiabatic speed of sound is ill-defined if $\phi'=0$. With this, we have expressed $w_{\phi}$, $c^2_{s\phi}$ and $c^2_{a\phi}$ in terms of the background quantities. Therefore, \eqref{originalperturbationeq1} and \eqref{originalperturbationeq2} are well defined for the scalar field and the multi-fluid approach is valid if $\phi'=0$ is not reached.

In addition to the evolution equations of $\delta_A$ and $v_A$ for each component $A$, we also have the perturbed Einstein equations, which read as \cite{baumann,Albarran:2016mdu}

\begin{equation}\label{perturbedeinsteineq1}
   \Psi'+\left(1+\frac{k^2}{3\mathcal{H}^2}\right)\Psi=-\frac{\delta}{2},
\end{equation}

\begin{equation}\label{perturbedeinsteineq2}
   \Psi'+\Psi=-\frac{3}{2}\mathcal{H}v(1+w_{\textrm{eff}}),
\end{equation}

\begin{equation}\label{perturbedeinsteineq3}
   \Psi''+\left[3-\frac{1}{2}(1+3w_{\textrm{eff}})\right]\Psi'-3w_{\textrm{eff}}\Psi=\frac{3}{2}\frac{\delta p}{\rho},
\end{equation}
where we have introduced the total fractional energy density perturbation, the total pressure perturbation, the total equation of state and the total peculiar velocity potential, which respectively read

\begin{equation}\label{totalperturbedquantities}
\begin{gathered}
    \delta=\sum_{A}\Omega_A\delta_A, \quad \delta p=\sum_{A}\delta p_A,\\ w_{\textrm{eff}}=\sum_{A}\Omega_Aw_A, \quad v=\sum_{A}\frac{1+w_A}{1+w_{\textrm{eff}}}\Omega_Av_A.
\end{gathered}
\end{equation}

We would like to clarify that all the formalism presented in this subsection is general and valid as long as the universe is described by a spatially flat FLRW metric. While the perturbations associated with the scalar field are often used to analyse the inflationary stage (in models where inflation is described by a scalar field), we have not made additional assumptions here, such as the slow-roll condition present in inflation. Of course, scalar field perturbations can be studied in an inflationary context, but this is not the purpose of this work. The formalism is equally valid for the analysis of the late-time universe.

\subsection{Adiabatic mode and comoving and homogeneous curvature perturbations}\label{admodandcurvpert}

In this subsection, we review the adiabatic mode as introduced by Weinberg in \cite{Weinberg:2003sw}, and also studied in \cite{Pajer:2017hmb}. This will be helpful to explain some of our results discussed in the following sections.

In the Newtonian gauge, in the limit of vanishing wave number, $k/\mathcal{H}\rightarrow0$, there is an extra gauge freedom given by a redefinition of the cosmic time and a rescaling of the spatial coordinate:

\begin{equation}\label{extragaugefreedom}
   t\rightarrow t+\epsilon(t), \quad x^i\rightarrow(1-\theta)x^i,
\end{equation}
where $\epsilon(t)$ is an arbitrary infinitesimal function of time and $\theta$ is an arbitrary infinitesimal constant. It can be shown that, in this limit, there is a physical solution which can be completely expressed in terms of the gauge transformation \cite{Weinberg:2003sw}:

\begin{equation}\label{gaugesolution}
\begin{gathered}
   \Psi=H\epsilon-\theta,\\ \delta\rho=-\epsilon\dot{\rho}, \quad \delta p=-\epsilon \dot{p}, \quad \delta\rho_A=-\epsilon\dot{\rho}_A, \quad \delta p_A=-\epsilon\dot{p}_A,\\ av=\epsilon, \quad av_A=\epsilon,
\end{gathered}
\end{equation}
where we have introduced the total background pressure, $p=\sum_Ap_A$. In the absence of an anisotropic stress tensor, as in our case, the gauge transformations $\epsilon$ and $\theta$ are related \cite{Weinberg:2003sw}:

\begin{equation}\label{epsilonlambdarelation}
   \epsilon(t)=\frac{\theta}{a(t)}\int^{t}a(t')dt',
\end{equation}
with an arbitrary lower limit. 

In a universe dominated by a fluid $B$ with constant equation of state $w_B$, the scale factor satisfies a power-law time dependence, $a\propto t^{2/[3(1+w_B)]}$, and the Hubble parameter is given by $H=\dot{a}/a=2/[3t(1+w_B)]$. With this, the dominant part of \eqref{epsilonlambdarelation} reads

\begin{equation}\label{dominantepsilon}
   \epsilon=\frac{2\theta}{H(5+3w_B)}.
\end{equation}
Taking into consideration this relation and the conservation equations for each component, $\dot{\rho}_A=-3H(1+w_A)\rho_A$, the physical solution given by \eqref{gaugesolution} can be rewritten as

\begin{equation}\label{gaugesolutionrewritten}
\begin{gathered}
   \Psi=-\frac{3\theta(1+w_B)}{5+3w_B},\\ \frac{\delta}{3(1+w_{\textrm{eff}})}=\frac{\delta_A}{3(1+w_A)}=\frac{2\theta}{5+3w_B},\\ \mathcal{H}v=\mathcal{H}v_A=\frac{2\theta}{5+3w_B},
\end{gathered}
\end{equation}
where we have eliminated the pressure relations since they are redundant \footnote{Note that, if $\delta\rho_A=-\epsilon\dot{\rho}_A$ and $av_A=\epsilon$, $\delta p_{naA}$=0, as can be deduced from \eqref{nonadpressure}. Therefore, \eqref{adandnonadpressure} simplifies to $\delta p_A=c^2_{aA}\delta\rho_{A}$, where $c^2_{aA}=\dot{p}_A/\dot{\rho}_A$ (cf. \eqref{speedsofsound}), and, since $\delta\rho_A=-\epsilon\dot{\rho}_A$, we find $\delta p_A=-\epsilon\dot{p}_A$.}.

On the other hand, the comoving curvature perturbation is defined as \cite{Bardeen:1980kt}

\begin{equation}\label{comovingcurvatureperturbation}
   \mathcal{R}=-\Psi+\mathcal{H}v.
\end{equation}
It can be shown that this mode is conserved in the super-Hubble regime $k/\mathcal{H}\ll1$ if and only if the total perturbation is adiabatic: $\delta\rho/\dot{\rho}=\delta p/\dot{p}$. Of course, the adiabatic mode satisfies this condition and $\mathcal{R}$ is conserved. In fact, from \eqref{gaugesolutionrewritten} we find that, in a universe dominated by a single fluid, the comoving curvature perturbation is given by $\mathcal{R}=\theta$. It is equally important to introduce the homogeneous curvature perturbation, which is defined as \cite{Bardeen:1983}

\begin{equation}\label{homogeneouscurvatureperturbation}
   \zeta=-\Psi-\delta\rho/\rho'.
\end{equation}
It is sometimes used instead of $\mathcal{R}$ since, in the super-Hubble limit $k/\mathcal{H}\rightarrow0$, the two perturbed quantities coincide \footnote{As noted in \cite{Gordon:2000hv}, $\delta\rho/\rho'\rightarrow -\mathcal{H}v$ in the limit $k/\mathcal{H}\rightarrow0$. Therefore, $\zeta\rightarrow \mathcal{R}$ in the limit $k/\mathcal{H}\rightarrow0$.}.

\subsection{Initial conditions}\label{inicond}

In order to solve numerically the perturbation equations, we must impose initial conditions. We assume that, initially, the Universe is dominated by the radiation component. We set the initial time at $\log{(a_i/a_0)=-15}$. Therefore, we have initially $\rho_i\approx\rho_{ri}$, $\delta_i\approx\delta_{ri}$, $\delta p_i\approx\delta p_{ri}$, $w_{\textrm{eff}i}\approx 1/3$ and $v_i\approx v_{ri}$. At this moment, all the relevant modes in the linear regime are super-Hubble and their wave number are much smaller than the comoving Hubble parameter, $k\ll\mathcal{H}$. Combining \eqref{perturbedeinsteineq1} and \eqref{perturbedeinsteineq3}, we obtain in this initial stage that the dominant solution is given by a constant metric potential, $\Psi\approx const.$ \cite{baumann,Albarran:2016mdu}. For this reason, we impose $\Psi'_i=0$. Therefore, from \eqref{perturbedeinsteineq1} and \eqref{perturbedeinsteineq2}, we deduce that initially the perturbations must satisfy

\begin{equation}\label{firstadinicond}
\begin{gathered}
   \Psi_i=-\frac{\delta_i}{2[1+k^2/(3\mathcal{H}_i^2)]},\\ \Psi_i=-\frac{3}{2}\mathcal{H}_iv_i(1+w_{\textrm{eff}i}).
\end{gathered}
\end{equation}
We have included the term $(k/\mathcal{H}_i)^2$ despite the fact that, initially, we are in the super-Hubble regime. This is necessary so that $\Psi'_i$ is exactly zero at the starting point of the simulations. If we eliminate this term, some small oscillations happen at the beginning, but a fast convergence to $\Psi'=0$ is reached. In addition, motivated by single scalar field models of inflation, we choose adiabatic initial conditions, which read as \cite{Albarran:2016mdu,Ballesteros:2010ks}

\begin{equation}\label{secondadinicond}
\begin{gathered}
   \frac{3\delta_{ri}}{4}=\delta_{mi}=\frac{\delta_{\phi i}}{1+w_{\phi i}}=\frac{\delta_i}{1+w_{\textrm{eff}i}},\\ v_{ri}=v_{mi}=v_{\phi i}=v_i.
\end{gathered}
\end{equation}
All together, from \eqref{firstadinicond} and \eqref{secondadinicond} we find that, initially, the perturbations must satisfy

\small
\begin{equation}\label{adiabaticinconditions}
\begin{gathered}
   \frac{3\delta_{ri}}{4}=\delta_{mi}=\frac{\delta_{\phi i}}{1+w_{\phi i}}=\frac{\delta_i}{1+w_{\textrm{eff}i}}=-\frac{2[1+k^2/(3\mathcal{H}_i^2)]\Psi_i}{1+w_{\textrm{eff}i}},\\ v_{ri}=v_{mi}=v_{\phi i}=v_i=-\frac{2\Psi_i}{3(1+w_{\textrm{eff}i})\mathcal{H}_i}.
\end{gathered}
\end{equation}
\normalsize
Note that the initial conditions are consistent with the relations in \eqref{gaugesolutionrewritten} in a universe initially dominated by radiation, $w_{\textrm{eff}i}\approx w_B=1/3$. In fact, the adiabatic mode is dominant until modes become sub-Hubble.

The comoving curvature perturbation \eqref{comovingcurvatureperturbation} is conserved for super-Hubble modes if the perturbation is adiabatic, $\delta\rho/\dot{\rho}=\delta p/\dot{p}$, as discussed in the previous subsection. Initially, since the Universe is dominated by radiation and $k\ll\mathcal{H}$, $\mathcal{R}$ is conserved. In fact, $\mathcal{R}$ is also conserved after the transition to matter until the mode enters the horizon. In the asymptotic future, when the Universe is dominated by the scalar field, the modes become super-Hubble again. Moreover, the perturbation also becomes adiabatic and $\mathcal{R}$ is conserved again, as we will see in detail in the next section.

As was done in \cite{Albarran:2016mdu}, we set $\Psi_i=1$ to perform the numerical calculations. After solving the equations, we rescale the solution by taking into account the physical value of $\delta_{\textrm{phys}}(k)$, whose value is inferred from the Planck observational fit to single field inflation \cite{Planck:2018vyg}:

\begin{equation}\label{physicalmode}
   \delta_{\textrm{phys}}(k)=\frac{2\pi}{3}\sqrt{2A_s}\left(\frac{k}{k_{\textrm{pivot}}}\right)^{\frac{n_s-1}{2}}k^{-3/2},
\end{equation}
where $A_s$ and $n_s$ are the scalar amplitude and spectral index of the primordial inflationary power spectrum, respectively,
corresponding to a pivot scale $k_{\textrm{pivot}}$. Here we use $\ln{(10^{10}A_s)}=3.0448$, $n_s=0.96605$ and $k_{\textrm{pivot}}=$ $0.05\,\,$Mpc$^{-1}$ \cite{Planck:2018vyg}.

\section{Numerical results}\label{numres}

In this section, we present the numerical results obtained when solving the perturbation equations discussed in the previous section. As we have shown, the equations for the scalar field perturbations depend on background quantities (cf. \eqref{fieldeos} and \eqref{scalarspeeds}). Therefore, to numerically solve the perturbation equations, we must first solve the dynamical system associated with the background. This system was numerically solved and analysed in our previous work \cite{Boiza:2024azh}. 

\subsection{Homogeneous curvature perturbation}\label{curvpert}

\begin{figure*}[t]
\centering
\begin{subfigure}{0.45\textwidth}
\includegraphics[width=\textwidth]{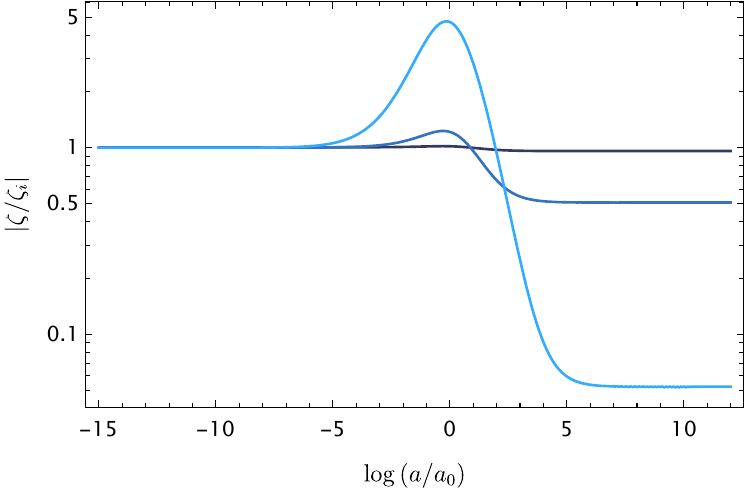}
\caption{Representation of the evolution of $\abs{\zeta/\zeta_i}$ for $n=1$.}
\label{fig:admod1}    
\end{subfigure}
\hspace{1cm}
\begin{subfigure}{0.45\textwidth}
\includegraphics[width=\textwidth]{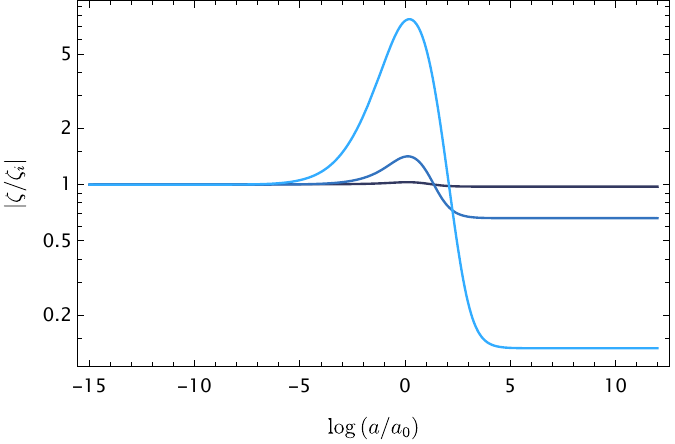}
\caption{Representation of the evolution of $\abs{\zeta/\zeta_i}$ for $n=0.1$.}
\label{fig:admod01}    
\end{subfigure}
\caption{\justifying{The left hand side panel shows the evolution of the absolute value of the homogeneous curvature perturbation normalised to its initial value, $\abs{\zeta/\zeta_i}$, for the case $n=1$. Each colour represents a different mode $k$ of the perturbations. We have represented three different modes: $k=10^{-4}$h Mpc$^{-1}$ (the last to enter the horizon, represented by dark blue), $k=4\times10^{-4}$h Mpc$^{-1}$ (represented by medium shade of blue) and $k=1.6\times10^{-3}$h Mpc$^{-1}$ (represented by light blue). Note that, for every mode represented, the linear regime is valid ($k<0.2$h Mpc$^{-1}$). The right hand side panel shows the evolution of $\abs{\zeta/\zeta_i}$ for the case $n=0.1$. Each colour represents the same mode as in the left hand side panel. Conservation of the homogeneous curvature perturbation is observed both at the beginning of the simulations, when adiabatic mode dominates, and at very late-time in the future, when the perturbation is in the super-Hubble regime and becomes adiabatic again.}}
\end{figure*}

In the previous section we introduced the comoving curvature perturbation, $\mathcal{R}$, and the homogeneous curvature perturbation, $\zeta$ (see \eqref{comovingcurvatureperturbation} and \eqref{homogeneouscurvatureperturbation}, respectively). In this work, we focus on $\zeta$, since the two curvature perturbations coincide in the super-Hubble limit $k/\mathcal{H}\rightarrow0$.

In Figs. \ref{fig:admod1} and \ref{fig:admod01}, we have represented the evolution of $\zeta$ for different modes in the cases $n=1$ and $n=0.1$, respectively. Initially, every mode is in the super-Hubble regime, $k/\mathcal{H}\ll1$, and the adiabatic mode dominates (cf. \eqref{gaugesolution}), so the homogeneous curvature perturbation is conserved. When the modes enter the horizon, the conservation is broken and some non-trivial evolution happens. We see that, in the far future, the homogeneous curvature perturbation is conserved again. In the final stage, the scalar field dominates and all of the modes exit the Hubble horizon and become super-Hubble again. $\mathcal{R}$ and $\zeta$ coincide in the super-Hubble limit because $\delta\rho/\rho'\rightarrow -\mathcal{H}v$ as $k/\mathcal{H}\rightarrow0$ \cite{Gordon:2000hv}. Since the scalar field dominates at late-time, we find from \eqref{nonadpressure} that the non-adiabatic part vanishes for super-Hubble modes and the perturbation becomes adiabatic \footnote{Note that, since in the final stage the scalar field dominates, we have $\rho\approx\rho_{\phi}$, $\delta\rho\approx\delta\rho_{\phi}$ and $v\approx v_{\phi}$. Since in the final stage all the modes in the linear regime are super-Hubble ($k/\mathcal{H}\ll1$), we also have $\delta\rho/\rho'\approx -\mathcal{H}v$. Therefore, $\delta\rho_{\phi}/\rho_{\phi}'\approx -\mathcal{H}v_{\phi}$ and the non-adiabatic part of the scalar field vanishes (see \eqref{nonadpressure} for $A=\phi$). It means that the total perturbation is adiabatic, since in this stage $\delta p\approx\delta p_{\phi}$.}. Therefore, $\zeta$ is conserved again. Note that the relations in \eqref{gaugesolution} are not generally satisfied in the final stage. Even so, the super-Hubble regime and the adiabaticity of the total perturbation guarantee the conservation of the homogeneous curvature perturbation, as in our case.

\subsection{Gravitational potential}\label{gravpot}

In the Newtonian gauge, in the absence of anisotropies, the scalar metric perturbations $\Psi$ and $\Phi$ introduced in \eqref{perturbedmetric} coincide: $\Psi=\Phi$. In this case, $\Psi$ plays the role of the gravitational potential. 

\begin{figure*}[t]
\centering
\begin{subfigure}{0.45\textwidth}
\includegraphics[width=\textwidth]{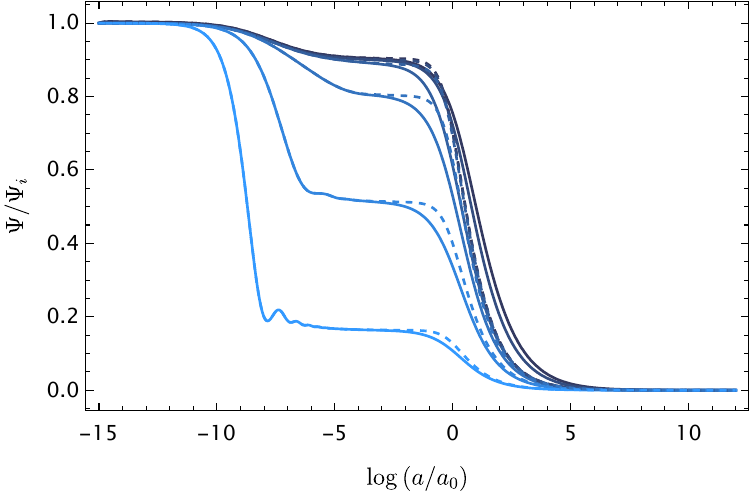}
\caption{Representation of the evolution of $\Psi/\Psi_i$ for $n=1$.}
\label{fig:gravpot1}    
\end{subfigure}
\hspace{1cm}
\begin{subfigure}{0.45\textwidth}
\includegraphics[width=\textwidth]{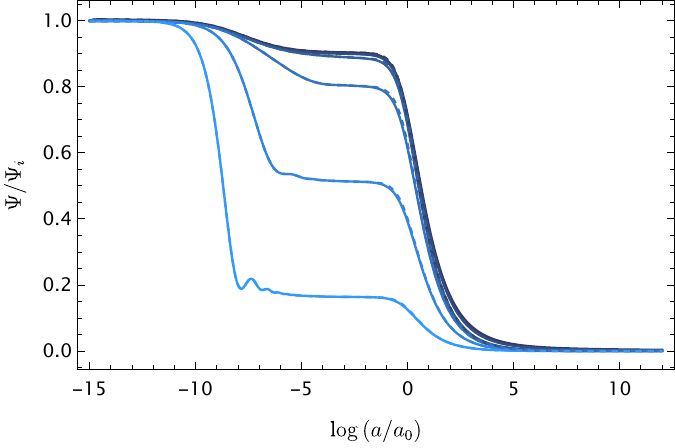}
\caption{Representation of the evolution of $\Psi/\Psi_i$ for $n=0.1$.}
\label{fig:gravpot01}    
\end{subfigure}
\caption{\justifying{The left hand side panel shows the evolution of the gravitational potential normalised to its initial value, $\Psi/\Psi_i$, for the case $n=1$. Each colour represents a different mode $k$ of the perturbations. We have represented six different modes ranging from $k=10^{-4}$h Mpc$^{-1}$ (the last to enter the horizon, represented by the darkest blue) to $k=0.1$h Mpc$^{-1}$ (the first to enter the horizon, represented by the lightest blue). Once again, we have selected each mode to ensure it falls within the validity of the linear regime ($k<0.2$h Mpc$^{-1}$). The solid lines represent the evolution of the gravitational potential in our model, while the dashed lines represent the evolution in $\Lambda$CDM for the same modes. The right hand side panel shows the evolution of $\Psi/\Psi_i$ for the case $n=0.1$. Each colour represents the same mode as in the left hand side panel. Bigger differences with respect to $\Lambda$CDM are observed for the $n=1$ case, when tracking happens far from $w_v=-1$.}}
\end{figure*}

In Figs. \ref{fig:gravpot1} and \ref{fig:gravpot01}, we have represented the evolution of the gravitational potential for different modes in the cases $n=1$ and $n=0.1$, respectively, and we have compared them with the evolution of $\Lambda$CDM. Initially, both cases have the same evolution and the differences with respect to a cosmological constant are not appreciable. Every mode is super-Hubble at the beginning of the simulations and the adiabatic mode dominates. Therefore, the gravitational potential remains constant and satisfies the relation given in \eqref{gaugesolutionrewritten}, with $w_B=1/3$. The modes that cross the horizon in the radiation-dominated era suffer from a strong suppression. In fact, these modes oscillate with frequency $k/\sqrt{3}$ and their amplitudes decrease as $a^{-2}$ in the sub-Hubble regime \cite{baumann}.

In the matter era, the dominant solution for the gravitational potential is a constant both in the super-Hubble and in the sub-Hubble regimes. Consequently, the modes that enter in the horizon in the matter era only suffer from a suppression due to the transition from radiation to matter. If during the transition the super-Hubble regime is still valid, the adiabatic mode is dominant and the gravitational potential satisfies the relation given in \eqref{gaugesolutionrewritten}, with $w_B=1/3$ in the radiation era and $w_B=0$ in the matter era. Therefore, the suppression factor is given by $\Psi_m/\Psi_r=9/10$, where $\Psi_m$ is the gravitational potential in the matter era and $\Psi_r$ is the gravitational potential in the radiation era. Apart from this suppression, the gravitational potential remains constant once the mode enters the horizon.

The differences with respect to $\Lambda$CDM are appreciable only at late-time, in the transition from matter to dark energy. For the case $n=1$, represented in Fig. \ref{fig:gravpot1}, we observe a suppression at low redshift with respect to $\Lambda$CDM. As mentioned in \cite{Boiza:2024azh}, in this case the matter component is suppressed in favour of the dark energy component due to tracking far from $w_v=-1$. From \eqref{perturbedeinsteineq2}, we see that $\Psi\rightarrow0$ and $\Psi'\rightarrow0$ asymptotically in the future when the potential of the scalar field dominates and $w_{\textrm{eff}}\rightarrow-1$. The transition to $\Psi\rightarrow0$ begins earlier if the scalar field acquires importance in the tracking epoch, as happens in this case, since $w_{\phi}$ is further from $-1$. We observe this effect in Fig. \ref{fig:gravpot1}. In the case $n=0.1$, represented in Fig. \ref{fig:gravpot01}, tracking occurs closer to $w_v=-1$ and the matter component does not suffer from a big suppression. In this case, the transition to $\Psi\rightarrow0$ at late-time is not accelerated and the differences with respect to $\Lambda$CDM are relaxed. 


\subsection{Dark energy perturbation}\label{depert}

In the previous section, we argued that the multi-fluid picture could be used in this work since we have no oscillatory regime of the scalar field in the cases analysed and $\phi'\neq0$. In this approach, the perturbations of the scalar field are studied by solving the evolution of $\delta\rho_{\phi}$ and $v_{\phi}$, defined in \eqref{scalardensityperturbation} and \eqref{scalarvelocityperturbation}, respectively. The scalar field behaves like a non-adiabatic fluid with $c^2_{s\phi}\neq c^2_{a\phi}$, both defined in \eqref{scalarspeeds}.

\begin{figure*}[t]
\centering
\begin{subfigure}{0.45\textwidth}
\includegraphics[width=\textwidth]{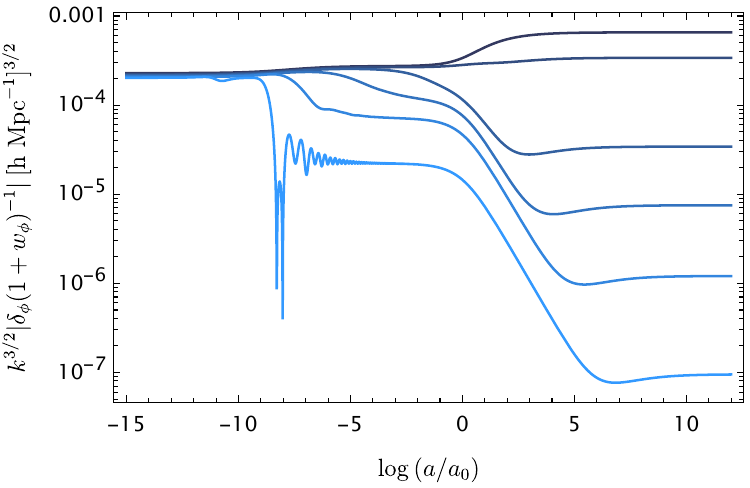}
\caption{Representation of the evolution of $k^{3/2}\abs{\delta_{\phi}(1+w_{\phi})^{-1}}$ for $n=1$.}
\label{fig:depert1}
\end{subfigure}
\hspace{1cm}
\begin{subfigure}{0.45\textwidth}
\includegraphics[width=\textwidth]{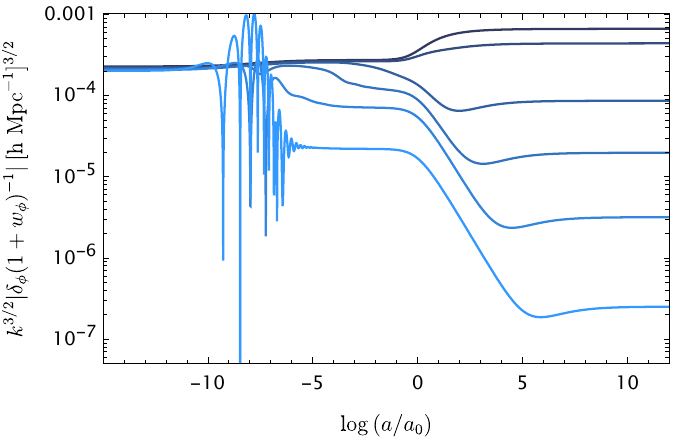}
\caption{Representation of the evolution of $k^{3/2}\abs{\delta_{\phi}(1+w_{\phi})^{-1}}$ for $n=0.1$.}
\label{fig:depert01}    
\end{subfigure}
\caption{\justifying{The left hand side panel shows the evolution of the scalar field perturbations through the quantity $k^{3/2}\abs{\delta_{\phi}(1+w_{\phi})^{-1}}$ for the case $n=1$. Each colour represents a different mode $k$ of the perturbations. The modes considered and the colour criterion are the same as for Figs. \ref{fig:gravpot1} and \ref{fig:gravpot01}. The right hand side panel shows the evolution of $k^{3/2}\abs{\delta_{\phi}(1+w_{\phi})^{-1}}$ for the case $n=0.1$. Each colour represents the same mode as in the left hand side panel.}}
\end{figure*}

In Figs. \ref{fig:depert1} and \ref{fig:depert01}, we have represented the evolution of $\delta_{\phi}/(1+w_{\phi})$ for different modes in the cases $n=1$ and $n=0.1$, respectively (we remind that $\delta_{\phi}=\delta\rho_{\phi}/\rho_{\phi}$). As previously mentioned, the adiabatic mode is the dominant one at the beginning, when every mode is super-Hubble and the Universe is dominated by radiation. In this early evolution, the relations \eqref{gaugesolutionrewritten} are satisfied with $w_B=1/3$ and we expect $\delta_{\phi}/(1+w_{\phi})$ to remain constant, as observed in the figures. The modes that do not enter the horizon during the radiation-dominated era experience an enhancement in the transition to matter-domination by a factor of $6/5$, since $w_B=0$ in the matter-dominated era. Once the modes enter the horizon, their non-adiabatic part becomes important and they decouple from the adiabatic mode. 

Every mode becomes super-Hubble again in the future and the total perturbation satisfies $\delta\rho/\rho'\approx-\mathcal{H}v$. Since the Universe is now dominated by the scalar field, this implies $\delta\rho_{\phi}/\rho'_{\phi}\approx-\mathcal{H}v_{\phi}$, as we discussed in Subsec. \ref{curvpert}. Therefore, the non-adiabatic part of the perturbation vanishes. After a Fourier transformation, $\nabla^2\rightarrow-k^2$, \eqref{originalperturbationeq1} can be rewritten as

\begin{equation}\label{originalperturbationeq1rewritten}
\begin{split}
   \left(\frac{\delta_A}{1+w_A}\right)'=\,\,&\left[9(c^2_{sA}-c^2_{aA})+\left(\frac{k}{\mathcal{H}}\right)^2\right]\mathcal{H}v_A\\
   &+3(c^2_{aA}-c^2_{sA})\frac{\delta_A}{1+w_A}+3\Psi'.
\end{split}
\end{equation}
The late-time regime is given by a universe dominated by the potential part of the scalar field ($w_{\phi}\approx-1$) in the super-Hubble limit ($k\ll\mathcal{H}$) and, therefore, $\Psi\approx0$, $\Psi'\approx0$, $\delta\rho_{\phi}/\rho'_{\phi}\approx -\mathcal{H}v_{\phi}$, as noted in Subsecs. \ref{curvpert} and \ref{gravpot}. From \eqref{originalperturbationeq1rewritten} applied to $A=\phi$, we find that \footnote{Note that, again, we have used the conservation equation for the scalar field: $\rho'_{\phi}+3(1+w_{\phi})\rho_{\phi}=0$. Therefore, $\delta_{\phi}/(1+w_{\phi})=-3\delta\rho_{\phi}/\rho'_{\phi}$ and, if $\Psi'\rightarrow0$ and $\delta\rho_{\phi}/\rho'_{\phi}\rightarrow -\mathcal{H}v_{\phi}$ in the super-Hubble limit ($k/\mathcal{H}\rightarrow0$), the right-hand side of \eqref{originalperturbationeq1rewritten} vanishes for $A=\phi$.}, at late-time, $\delta_{\phi}/(1+w_{\phi})\approx const.$ We observe this behaviour in Figs. \ref{fig:depert1} and \ref{fig:depert01}. Since $w_{\phi}$ asymptotically decays to $-1$ in the cases analysed, we deduce that $\delta_{\phi}$ must also decay asymptotically to $0$ with a rate similar to $\dot{w}_{\phi}$.

\section{Cosmological observables}\label{cosobs}

After the study conducted in the previous section on perturbations in our quintessence model, we now focus on structure formation in this section. We will explore how observables such as the matter power spectrum and the $f\sigma_8$ distribution can impose constraints on the parameters of the models.

\begin{figure*}[t]
\centering
\begin{subfigure}{0.45\textwidth}
\includegraphics[width=\textwidth]{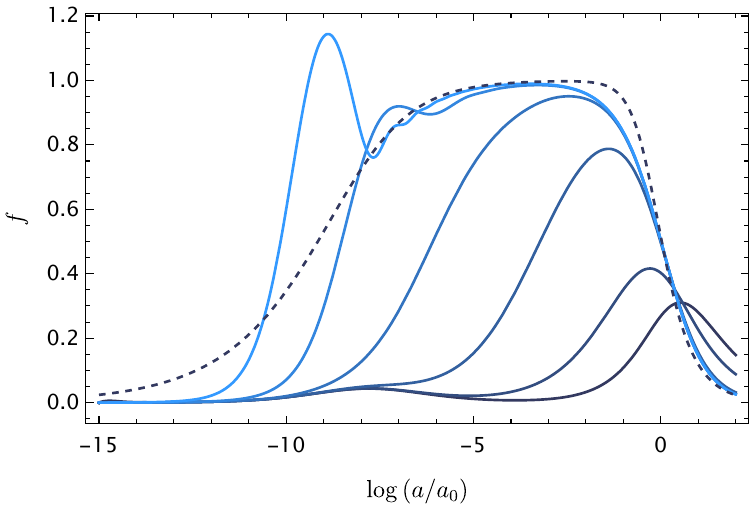}
\caption{Representation of the evolution of $f$ for $n=1$.}
\label{fig:growthfunction1}    
\end{subfigure}
\hspace{1cm}
\begin{subfigure}{0.45\textwidth}
\includegraphics[width=\textwidth]{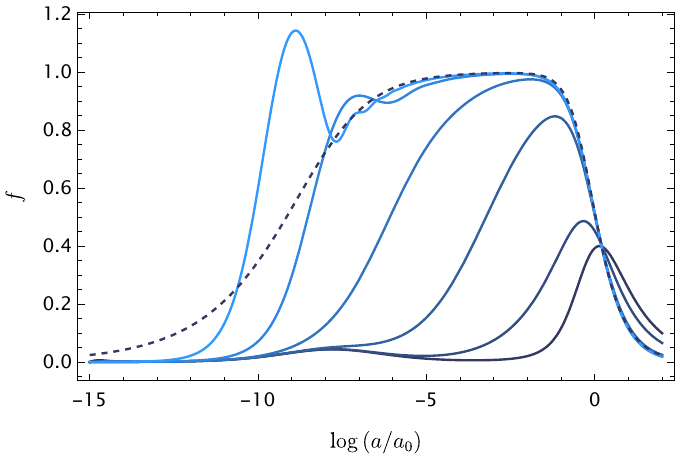}
\caption{Representation of the evolution of $f$ for $n=0.1$.}
\label{fig:growthfunction01}    
\end{subfigure}
\caption{\justifying{The left hand side panel shows the evolution of the growth rate function $f$ for $n=1$. Each colour represents a different mode $k$ of the perturbations. The modes considered and the colour criterion are the same as for Figs. \ref{fig:gravpot1} and \ref{fig:gravpot01}. The solid lines represent the evolution of the growth rate function in our model, while the dashed line represents the approximation $f_{\Lambda}\approx\Omega^{\gamma}_{\Lambda m}$ for $\Lambda$CDM, valid in the sub-Hubble regime ($k\gg\mathcal{H}$). The right hand side panel shows the evolution of $f$ for $n=0.1$. Each colour represents the same mode as in the left hand side panel. Higher suppression with respect to $\Lambda$CDM is observed for the $n=1$ case.}}
\end{figure*}

In Figs. \ref{fig:growthfunction1} and \ref{fig:growthfunction01}, we have represented the evolution of the growth rate function, $f=\delta'_m/\delta_m$ \cite{Balcerzak:2012ae,Albarran:2016mdu}, for different modes in our model. We compare them with the $\Lambda$CDM estimation, $f_{\Lambda}\approx\Omega^{\gamma}_{\Lambda m}$, where $\gamma=0.55$ \cite{Wang:1998gt,Linder:2005in,Linder:2007hg}, valid in the sub-Hubble regime $k\gg\mathcal{H}$. Initially, all the modes are super-Hubble and matter perturbations remain almost constant. Therefore, $f\approx0$ until the modes enter the horizon. Then, perturbations grow and their growth rates reach a common path given by the approximate function $f\approx\Omega^{\gamma}_{m}$, where $\gamma$ is now given by $\gamma=0.55+0.05[1+w_{\phi}(z=1)]$ \cite{Linder:2005in}. Note that this parametrisation is a correction for general dark energy models: we recover $\gamma=0.55$ in $\Lambda$CDM since $w_v=-1$ at every redshift. In the sub-Hubble regime, $k$ dependence is lost in the growth function and we can make a comparison between our model predictions and $\Lambda$CDM ones. For the case of $n=1$ represented in Fig. \ref{fig:growthfunction1}, we observe a suppression with respect $\Lambda$CDM at low redshift. As explained in the previous subsection, tracking with $n=1$ implies an equation of state $w_{\phi}$ further from $-1$. Therefore, matter component is suppressed in favour of dark energy. The same effect was observed for $\Omega_m$ in \cite{Boiza:2024azh}. Note that, in fact, the growth rate of sub-Hubble modes are well approximated by a parametrisation which depends on $\Omega_m$: if we found a suppression in $\Omega_m$, it is expected to find it also in $f$. For the case of $n=0.1$ represented in Fig. \ref{fig:growthfunction01}, sub-Hubble modes fit well to the $\Lambda$CDM approximation at lower redshift. In this case, $w_{\phi}$ in the tracking regime is closer to $-1$ and $\Omega_m$ is less suppressed \cite{Boiza:2024azh}. Therefore, the differences in $f$ with respect $\Lambda$CDM are less appreciable.    

The matter power spectrum gives us information on the distribution of matter over various scales in the Universe. In the Newtonian gauge that we have used in this work, it can be expressed as \cite{Albarran:2016mdu}

\begin{equation}\label{mps}
   P=\abs{\delta_m-3\mathcal{H}v_m}^2.
\end{equation}
Another important quantity which usually appears in the observational data when studying structure formation is $f\sigma_8$, a combination of the growth structure function $f$ previously represented in Figs. \ref{fig:growthfunction1} and \ref{fig:growthfunction01} and $\sigma_8$, which measures the root mean square (rms) fluctuation of the matter density field on a scale of 8h$^{-1}$ Mpc. The temporal evolution of $\sigma_8$ is given by \cite{Wang:2010gq,Albarran:2016mdu}

\begin{equation}\label{sigma8}
   \sigma_8(z,k_{\sigma_8})=\sigma_8(0,k_{\sigma_8})\frac{\delta_m(z,k_{\sigma_8})}{\delta_m(0,k_{\sigma_8})},
\end{equation}
where $k_{\sigma_8}=0.125$h Mpc$^{-1}$ is the wave-length of the mode corresponding to distances of 8h$^{-1}$ Mpc and we use $\sigma_8(0,k_{\sigma_8})=0.8120$ \cite{Planck:2018vyg}.

\begin{figure*}[t]
\centering
\begin{subfigure}{0.45\textwidth}
\includegraphics[width=\textwidth]{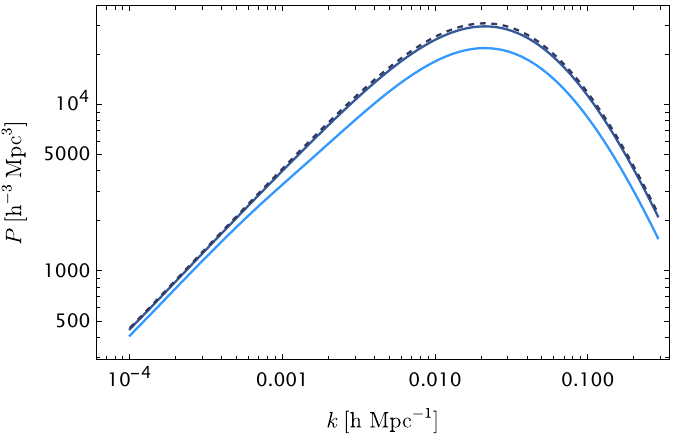}
\caption{Representation of matter power spectrum for $n=1$ and $n=0.1$.}
\label{fig:mpsall}  
\end{subfigure}
\hspace{1cm}
\begin{subfigure}{0.43\textwidth}
\includegraphics[width=\textwidth]{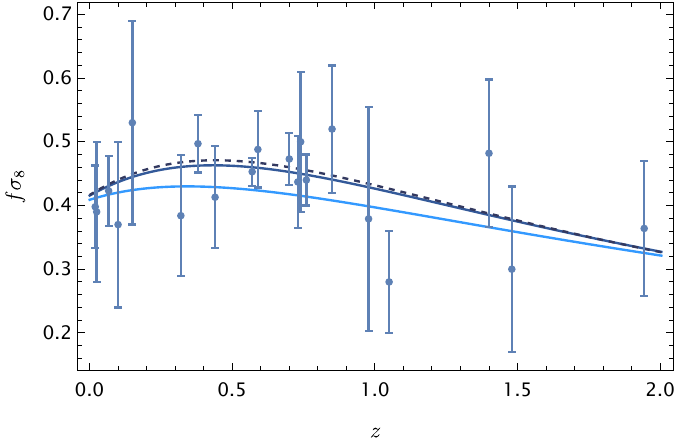}
\caption{Representation of the evolution of $f\sigma_8$ for $n=1$ and $n=0.1$.}
\label{fig:fs8all}   
\end{subfigure}
\caption{\justifying{The left panel shows the matter power spectrum from $k=10^{-4}$h Mpc$^{-1}$ to $k=0.3$h Mpc$^{-1}$. Note that the linear regime is valid till $k=0.2$h Mpc$^{-1}$; larger values of $k$ have been added in order to better observe the differences of the model with respect to $\Lambda$CDM in the high $k$ limit. The solid lines represent the numerical results of the model: the light blue for $n=1$ and the dark blue for $n=0.1$. The dashed line represents the $\Lambda$CDM estimations. Higher suppression with respect to $\Lambda$CDM is observed for the $n=1$ case. The right panel shows the evolution of $f\sigma_8$ from redshift $z=2$ to $z=0$. The same criterion is used: the solid lines represent the numerical results of the model (the light blue for $n=1$ and the dark blue for $n=0.1$), while the dashed line represents the $\Lambda$CDM estimations. Observational data have been included in the plot \cite{Avila:2022xad}. Again, higher suppression with respect to $\Lambda$CDM is observed for the $n=1$ case.}}
\end{figure*}

In Fig. \ref{fig:mpsall}, we have represented the matter power spectrum for $n=1$ and $n=0.1$ at $z=0$ and we have compared with respect to $\Lambda$CDM. As expected, further suppression with respect to $\Lambda$CDM is found for $n=1$, as tracking occurs far from $w_v=-1$. The matter component is suppressed and, as a consequence, structure formation is disfavoured. For $n=0.1$, the difference with respect to $\Lambda$CDM is less appreciable. Tracking happens close to $w_v=-1$ and the matter component evolution mimics the $\Lambda$CDM one. Therefore, structure formation is larger at lower redshift and the suppression of the matter power spectrum is relaxed. In Fig. \ref{fig:fs8all}, we have represented the temporal evolution of $f\sigma_8$ for $n=1$ and $n=0.1$ in the low redshift range $z\in[0,2]$. We find the same suppression that was present in Figs. \ref{fig:growthfunction1} and \ref{fig:growthfunction01}: for $n=1$, the deviation from $\Lambda$CDM is bigger than for $n=0.1$.

From this section, we find that the structure formation is sensitive to the parameters of the theory. Specifically, having fixed $\eta$ to the reduced Planck mass, $\eta=M_P$, structure formation imposes extra constraints in the parameters. Throughout this work, we have deeply analysed the cases $n=1$ and $n=0.1$. We found that both scenarios allow for late-time acceleration in a universe dominated by the dark energy sector today: $\Omega_{m0}\approx0.3$ and $\Omega_{\phi0}\approx0.7$. Observational data on matter structures can help us to further constrain the values of the parameters. The complete evolution of the scalar field plays an important role, since tracking behaviour allows for scenarios in which the scalar field is not negligible at low redshift and can have appreciable effects in the matter structures which we observe today. In cases in which we are leaving the tracking regime nowadays, as the cases studied here, deviating so much from $w_v=-1$ can be problematic since we expect big suppression of matter structures with respect to $\Lambda$CDM. We have seen this effect by considering the specific cases $n=1$ and $n=0.1$.

\section{Conclusions}\label{conclusions}

In this work, we first reviewed the generalised axion-like model proposed in \cite{Boiza:2024azh}. This model arises as a generalisation of the axion-like potential to negative powers (cf. \eqref{axionscalarpotential}). We have reviewed the most important dynamical aspects of the model, which we discussed in depth in \cite{Boiza:2024azh}. Specifically, we have shown how the model can explain the late-time acceleration of the Universe through an effective cosmological constant that comes into play when the scalar field reaches the non-vanishing minimum of the potential (cf. \eqref{minimumexpansion}). We have also reviewed how the model exhibits tracking behaviour, which allows for an alleviation of the coincidence problem. Additionally, we have shown that in our model, the transition from matter to dark energy is favoured. This is because $\Gamma$ (see \eqref{gammadef} for a general definition of $\Gamma$ and \eqref{gammamodel} for the expression that $\Gamma$ takes in our model) becomes very large when the scalar field reaches the de Sitter-like attractor at $\lambda=0$. As a result, the amount of dark energy increases sharply in the final phase of the tracking regime (cf. \eqref{growgamma}). In \cite{Hossain:2023lxs}, the authors noted that this mechanism distinguishes the generalised axion-like model from an inverse power-law potential. In the former, the equation of state today is closer to $-1$.

After presenting and reviewing the dynamical aspects of the model, we examined the perturbation theory in the linear regime within the Newtonian gauge. We showed that the multi-fluid picture used in \cite{Albarran:2016mdu} is applicable to our quintessence model, as the considered cases ($\eta=M_P, \, n=1 \:\: \& \:\: \eta=M_P, \, n=0.1$) do not lead to an oscillatory regime for the scalar field; in other words, we avoid points where $\phi'=0$, at least up to the calculated times. Therefore, the peculiar velocity \eqref{scalarvelocityperturbation} and the adiabatic speed of sound \eqref{scalarspeeds} are well-defined. In the perturbed scalar field equations, some quantities depend on the dynamical evolution of the background (cf. \eqref{fieldeos} and \eqref{scalarspeeds}). Thus, to solve the perturbation equations, we must first solve the background dynamical system, which we have already described in \cite{Boiza:2024azh}. 

We have also analysed the adiabatic mode in Subsec. \ref{admodandcurvpert}, first proposed by Weinberg in \cite{Weinberg:2003sw}, and the comoving and homogeneous curvature perturbations (see \eqref{comovingcurvatureperturbation} and \eqref{homogeneouscurvatureperturbation}), which are conserved in the super-Hubble regime, $k\ll\mathcal{H}$, when the total perturbation is adiabatic. To perform a numerical calculation of the perturbations and observe their evolution, we have imposed adiabatic initial conditions \eqref{adiabaticinconditions}, motivated by single-field inflationary models.

After the theoretical introduction to perturbation theory applied to a quintessence model and after selecting the initial conditions, we conducted a study of the perturbations by numerically solving the equations. This study was carried out for two cases: $\eta=M_P, \, n=1 \:\: \& \:\: \eta=M_P, \, n=0.1$. First, we have analysed the evolution of the homogeneous curvature perturbation \eqref{homogeneouscurvatureperturbation}, which in the super-Hubble regime, $k\ll\mathcal{H}$, coincides with the comoving curvature perturbation \eqref{comovingcurvatureperturbation}. We found that the homogeneous curvature perturbation is conserved at the beginning of the evolution, when we are in the super-Hubble regime and the adiabatic mode \eqref{gaugesolution} dominates. When the modes enter the Hubble horizon, the homogeneous curvature perturbation ceases to be conserved and exhibits a non-trivial evolution. In the future, the Universe is dominated by the scalar field, and the total perturbation becomes adiabatic again. Since the super-Hubble regime is valid once more in this phase, we find that the homogeneous curvature perturbation is conserved again. It is worth noting that, although the total perturbation becomes adiabatic, the adiabatic mode introduced in Subsec. \ref{admodandcurvpert} is not recovered in this final stage.

We have also studied the evolution of the gravitational potential for the two cases mentioned above. In both cases, we observe similar behaviour at the beginning. When in the super-Hubble regime and the adiabatic mode dominates, the gravitational potential remains constant. Modes that stay in the super-Hubble regime during the radiation-to-matter transition experience a suppression by a factor of $9/10$. Those modes that enter the Hubble horizon before the transition suffer a greater suppression and subsequent oscillation, with a frequency $k/\sqrt{3}$. In the final stage of the evolution, during the matter-to-dark energy transition, the differences between the two cases become apparent. As discussed in \cite{Boiza:2024azh}, the amount of matter undergoes a larger suppression for higher values of the parameter $n$. Thus, we observe that the transition of the gravitational potential toward $\Psi\rightarrow0$, $\Psi'\rightarrow0$, begins earlier in the case of $n=1$. For the case $n=0.1$, the difference from $\Lambda$CDM is nearly imperceptible.

We have also analysed the evolution of dark energy perturbations. We found that, initially, when the adiabatic mode dominates and we are in the super-Hubble regime, the quantity $\delta_{\phi}/(1+w_{\phi})$ remains constant, as expected. When the modes enter the Hubble horizon and the super-Hubble regime is lost, the modes decouple from the adiabatic mode and exhibit non-trivial evolution. By the end of the evolution, we find that the total perturbation, which is approximately given by the dark energy perturbation (since the scalar field dominates), becomes adiabatic again. Since the super-Hubble regime is restored in this final phase, the quantity $\delta_{\phi}/(1+w_{\phi})$ is conserved once more. Thus, $\delta_{\phi}$ asymptotically decays toward $0$ at the same rate as $\dot{w}_{\phi}$. 

Finally, we have studied the structure formation for the two cases $n=1$ and $n=0.1$. The suppression of the parameter $\Omega_m$ that we found in \cite{Boiza:2024azh}, which is greater for higher values of $n$, translates into a suppression in structure formation as compared to $\Lambda$CDM. We have observed that in the case of $n=1$, the suppression of the growth rate function is larger than in the case of $n=0.1$. We found the same behaviour when studying the matter power spectrum and the $f\sigma_8$ distribution. Therefore, the data we have on structure formation could constrain the values of the parameter $n$ in our model.

To conclude this work, we note that variations in the parameter $\eta$ could also affect the evolution of the Universe. In fact, values smaller than the reduced Planck mass, $\eta<M_P$, may lead to an oscillatory regime where the perturbations cannot be studied within the multi-fluid framework, as $\phi'=0$ (see Subsec. \ref{frame}). We will conduct a detailed study of this case in an upcoming work \cite{cgbmariam}. We will also constrain the parameter space by performing a numerical fit to cosmological observations in \cite{hsuwencgbmariam}.

\section*{Acknowledgements}

The authors are grateful to Hsu-Wen Chiang, Jose Beltrán Jiménez, Florencia A. Teppa Pannia, Nelson J. Nunes and Tom Broadhurst for discussions and insights on the current project. C. G. B. acknowledges financial support from the FPI fellowship PRE2021-100340 of the Spanish Ministry of Science, Innovation and Universities. M. B.-L. is supported by the Basque Foundation of Science Ikerbasque. Our work is supported by the Spanish Grants PID2020-114035GB-100 (MINECO/AEI/FEDER, UE) and PID2023-149016NB-I00 (MINECO/AEI/FEDER, UE). This work is also supported by the Basque government Grant No. IT1628-22 (Spain). 

\bibliography{bibliografia.bib}

\end{document}